\documentclass[aps,prl,twocolumn,letterpaper,superscriptaddress,showpacs]{revtex4}

\usepackage{amsmath, amsthm}
\usepackage{keyval}
\usepackage{graphicx,latexsym}
\usepackage{dcolumn}
\usepackage{bm}
\usepackage{color}
\usepackage{CJK}

\begin{document}
\title{Transition between two metallic ferroelectric orders in multiferroic Ca$_3$Ru$_2$O$_7$, induced by magnetism-mediated orbital re-polarization}

\begin{CJK*}{UTF8}{}

\author{Zheting Jin (\CJKfamily{gbsn}金哲霆)}
\affiliation{Tsung-Dao Lee Institute \& School of Physics and Astronomy, Shanghai Jiao Tong University, Shanghai 200240, China}

\author{Wei Ku (\CJKfamily{bsmi}顧威)}
\altaffiliation{corresponding email: weiku@mailaps.org}
\affiliation{Tsung-Dao Lee Institute \& School of Physics and Astronomy, Shanghai Jiao Tong University, Shanghai 200240, China}
\affiliation{Key Laboratory of Artificial Structures and Quantum Control (Ministry of Education), Shanghai 200240, China}

\date{\today}

\begin{abstract}
For the past decades, the low-temperature phase of Ca$_3$Ru$_2$O$_7$ below the 48K first-order phase transition remains a puzzle with controversial suggestions involving metallic ferroelectric, orbital or magnetic ordering.
Through analysis of experimental lattice structure, density functional theory calculation, and effective model analysis, we propose that the 48K phase transition is a bond formation transition promoted by the magnetism mediated orbital re-polarization.
Most interestingly, this transition is accompanied by a switch of \textit{two} metallic ferroelectric orders from a $xy+y$ symmetry to $xz+z$.
Our study not only resolves a long-standing puzzle of this phase transition in this material, but also demonstrates perhaps the first example of transition between multiple emergent ferroelectric orders in bad metals, resulting from interplay between multiferroic orders.
\end{abstract}

\maketitle
\end{CJK*}

The Ruddlesden-Popper series of ruthenates ((Sr,Ca)$_{n+1}$Ru$_{n}$O$_{3n+1}$), consisted of layered corner-sharing RuO$_6$ octahedra, are interesting due to the coexistence of various electronic and magnetic orders, such as superconductivity~\cite{kidwingira2006dynamical}, Mott metal-insulator transition~\cite{braden1998crystal,puchkov1998layered,paper22}, orbital ordering~\cite{paper31}, meta-magnetism~\cite{paper30,paper21}, and unstable ferromagnetism~\cite{ikeda2000ground}.
The slightly more extended 4$d$ orbitals and their stronger spin-orbital couplings~\cite{paper19} in these materials further enrich the physics in comparison with the extensively studied 3$d$ transition-metal oxides.
For example, while CaRuO$_{3}$ is a paramagnetic metal~\cite{cao1997thermal,longo1968magnetic}, the isovalent SrRuO$_{3}$ is a ferromagnetic metal~\cite{allen1996transport} with a Fermi liquid ground state~\cite{mackenzie1998observation}.
Similarly, while Ca$_{2}$RuO$_{4}$ is a typical Mott insulator~\cite{paper31}, the isovalent Sr$_{2}$RuO$_{4}$ turns out to be an unconventional superconductor~\cite{kidwingira2006dynamical} possibly with a triplet pairing~\cite{mackenzie2003superconductivity}.

Specifically, in Ca$_3$Ru$_2$O$_7$, it has been suggested to host orbital order~\cite{paper8,paper5,paper9,paper4,cao2004orbitally}, metallic ferroelectric order~\cite{paper25}, and metallic antiferromagnetic order at $T_{N}=56$K~\cite{cao1997observation, paper9, paper29}.
Furthermore, a first-order resistivity-jumping transition is found at $T_{MI}=48$K~\cite{cao1997observation}, accompanied by a crystal collapse in $c$ direction.
The physics origin of this transition, however, remains controversial to date.

Several pictures have been proposed to explain this transition.
First, the transition was interpreted as a Mott insulator transition, based on the greatly enhanced resistivity at low temperature~\cite{cao1997observation,paper6,paper7} and Raman-scattering study~\cite{liu1999raman}. 
However, more recent resistivity studies on cleaner samples found metallic behavior at low temperature below 30K~\cite{paper21,paper30,paper28}, and only in some slightly doped samples this transition gives largely enhanced resistivity by as large as 8 orders of magnitude~\cite{paper12,paper27,paper32}.
Note however, that even for the doped samples, no exponential growth of the resistivity was ever reported to indicate a well-defined energy scale of the insulating gap, so the classification through a Mott insulator is not satisfactory.

Second, inspired by the $c$-axis lattice collapse at this transition~\cite{paper21}, a picture of orbital order with a increased occupation of $d_{xy}$ was proposed~\cite{paper10} to explain this transition.
This was, however, disproved later by a series of structure analyses, which showed that the dominant reason of this collapse is the larger inter-planar tilting of the octahedra~\cite{paper6,paper18,paper5,paper24}, without obvious shortening of the octahedra height.

More recently, at $T=9$K, angular resolved photoemission spectroscopy (ARPES) study~\cite{paper3} found diamond-shaped low-energy gapped excitations near the chemical potential around $(\pi,0)$ and $(0,\pi)$.
Assuming that the gapped excitations results from the gapping of a well nested diamond-shaped Fermi surface below 48K, a finite momentum $q\sim(\pi,\pi)$ order (eg: charge density wave or spin density wave) was proposed as a possible candidate for the low-temperature phase.
However, no additional Bragg peak near $(\pi,\pi)$ was ever seen in the diffraction studies~\cite{paper4,paper29,paper9}, casting doubt on this scenario.

This long-standing puzzle thus leaves many important questions that require a timely resolution:
1) What is the phase below 48K?  Or, what is the nature of this first order transition,
2) Why is the $c$-axis octahedral tilting enhanced in this phase?
3) Is there orbital order in this system?
4) Is there metallic ferroelectric order in this system?
5) Is there a strong interplay between this phase and the magnetic ordering to promotes the close proximity of their transition temperatures?
6) What is the interplay between the multiple orders?

\begin{figure}[htp!]
\begin{center}
\includegraphics[width=\columnwidth,clip,angle=0]{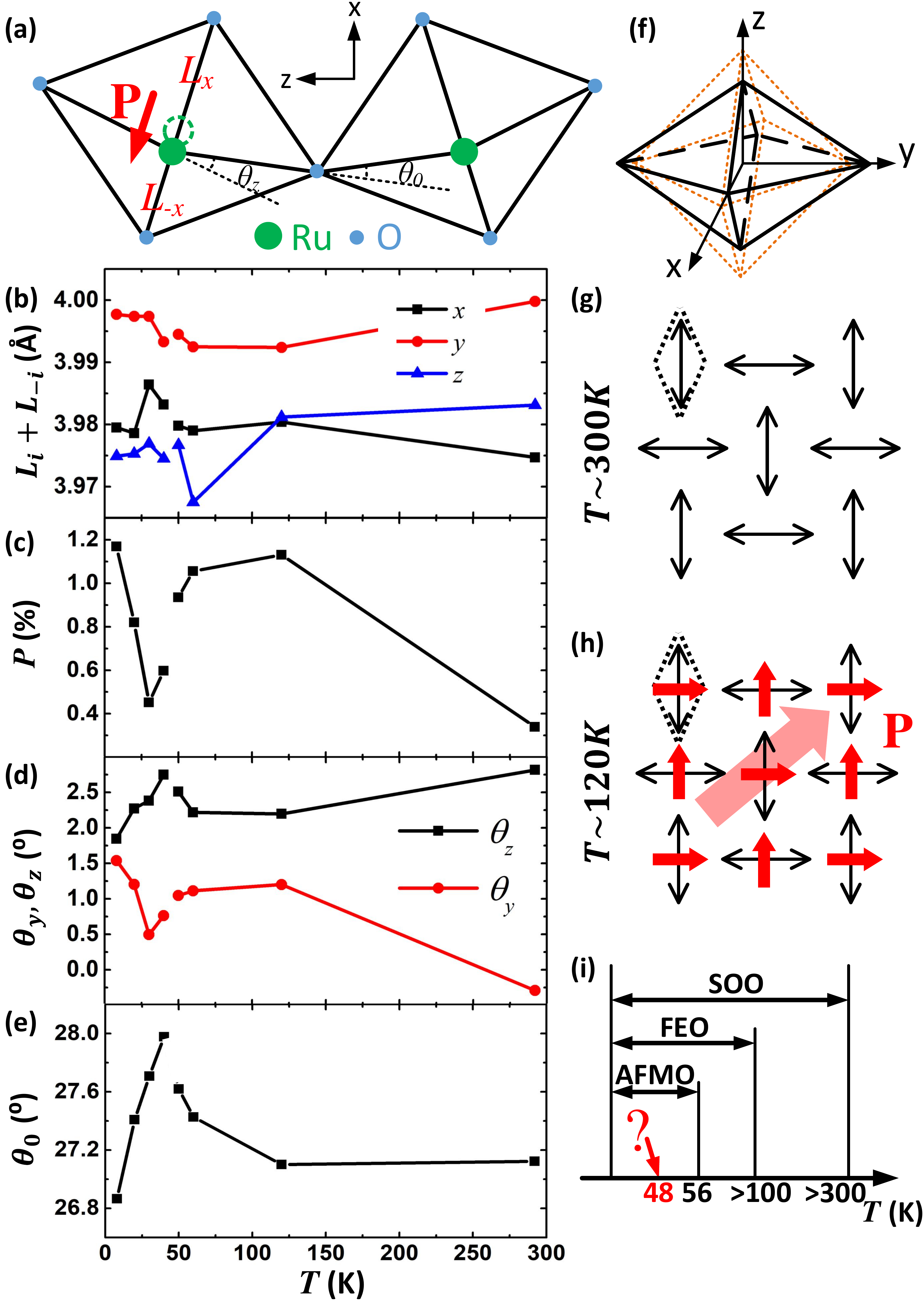}
\end{center}
\vspace{-0.8cm}
\caption{
\label{fig:fig1}
(color online) Summary of observed structure.
(a) Illustration of the bilayer octahedra, showing FEO related deviation of Ru position.
(b) Octahedral bond length $L_i+L_{-i}$, ($i=x,y,z$) in major axises, whose difference reflects orbital order.
(c) FEO order parameter $P=\frac {L_x-L_{-x}}{L_x+L_{-x}}$ from ochthedral bond length.
(d) O-Ru-O band angle, $\theta _z$, along $z$ axis and $\theta _y$ along $y$ axis.
(e) Octahedral tilt angles, $\theta _0$, defined in (a).
(f) Octahedral elongation and (g) SOO pattern at $T\sim 300$K.
(h) FEO pattern at $T\sim 120$K.
(i) Illustration of multiple orderings.
}
\vspace{-0.5cm}
\end{figure}

In this letter, we address these outstanding questions via a combined theoretical and computational study.
From careful analysis of experimental lattice structure and density functional theory (DFT) calculation, we found that orbitals indeed order in this system, but at very high temperature $T>300$K.
Similarly, a metallic ferroelectric order also emerges between 300K and 100K.
Under these two orders, we then found two distinct stable solutions that resemble experimental feature of the states above and below the 48K phases.
We then verify these two states through a LDA+$U$+$V$ model against our DFT results quantitatively.
These restuls suggest a scenario of first-order transition via inter-planar bonding induced by a magnetism-mediated orbital re-polarization.
Most interestingly, the strengthened bonding of the low-temperature phase switches the metallic ferroelectric order from $xy+y$ to a \textit{distinct} $xz+z$ order.
Our study not only resolves a long-standing puzzle of the 48K phase transition in this material, but also demonstrates perhaps the first example of transition between multiple emergent ferroelectric orders in bad metals, resulting from interplay between multiferroic orders.

First, we conclude a staggered orbital order (SOO) in this system using Fig~\ref{fig:fig1} that summarizes our analyses of existing experimental lattice structure~\cite{paper29}.
One notices from Fig.~\ref{fig:fig1}(b)(f) that even at the highest temperature being measured ($\sim$300K), the local $y$-axis of the distorted octahedra is already apparently longer than the $x$- and $z$-axis.
In the crystal, this longer octahedron axis orients at the global $x$ and $y$ directions in a staggered pattern in the $xy$ plane [cf: Fig.~\ref{fig:fig1}(g)].
The spontaneous lowering of symmetry related to $x\ne y$ in the local octahedra of such a bilayer system defines a thermodynamic order isomorphic to orbital order with different electronic occupation of the originally symmetric $xz$ and $yz$ orbitals in $t_{2g}$-active systems.
That is, the existing experimental structural data indicates clearly that an in-plane SOO has already taken place above the room temperature.

Next, we conclude a metallic ferroelectric order (FEO) in this system using Fig~\ref{fig:fig1}(c), which gives the local ferroelectric order parameter, defined via the local deviation of Ru atom from the center of the octahedron along the local $x$-direction, $P=\frac {L_x-L_{-x}}{L_x+L_{-x}}$.
It shows that a FEO (or more precisely a ferri-electric order) actually develops between 120 and 300K, and already becomes quite strong by 120K, with ferroelectric moment along the (1,1,0) directions and a staggered electric dipole moment along the (1,-1,0) direction [cf: Fig.~\ref{fig:fig1}(h)].
It is interesting to notice that the deviation of Ru position from the center of the octahedra (local electric dipole) is perpendicular to the local long-axis of the octahedron.
This seems quite counterintuitive from classical pictures, since moving atoms toward emptier space should have lower energy.
As we will explain below, this characteristic in fact reflects the microscopic electronic origin of this particular order.

\begin{figure}[th]
\begin{center}
\includegraphics[width=\columnwidth,clip,angle=0]{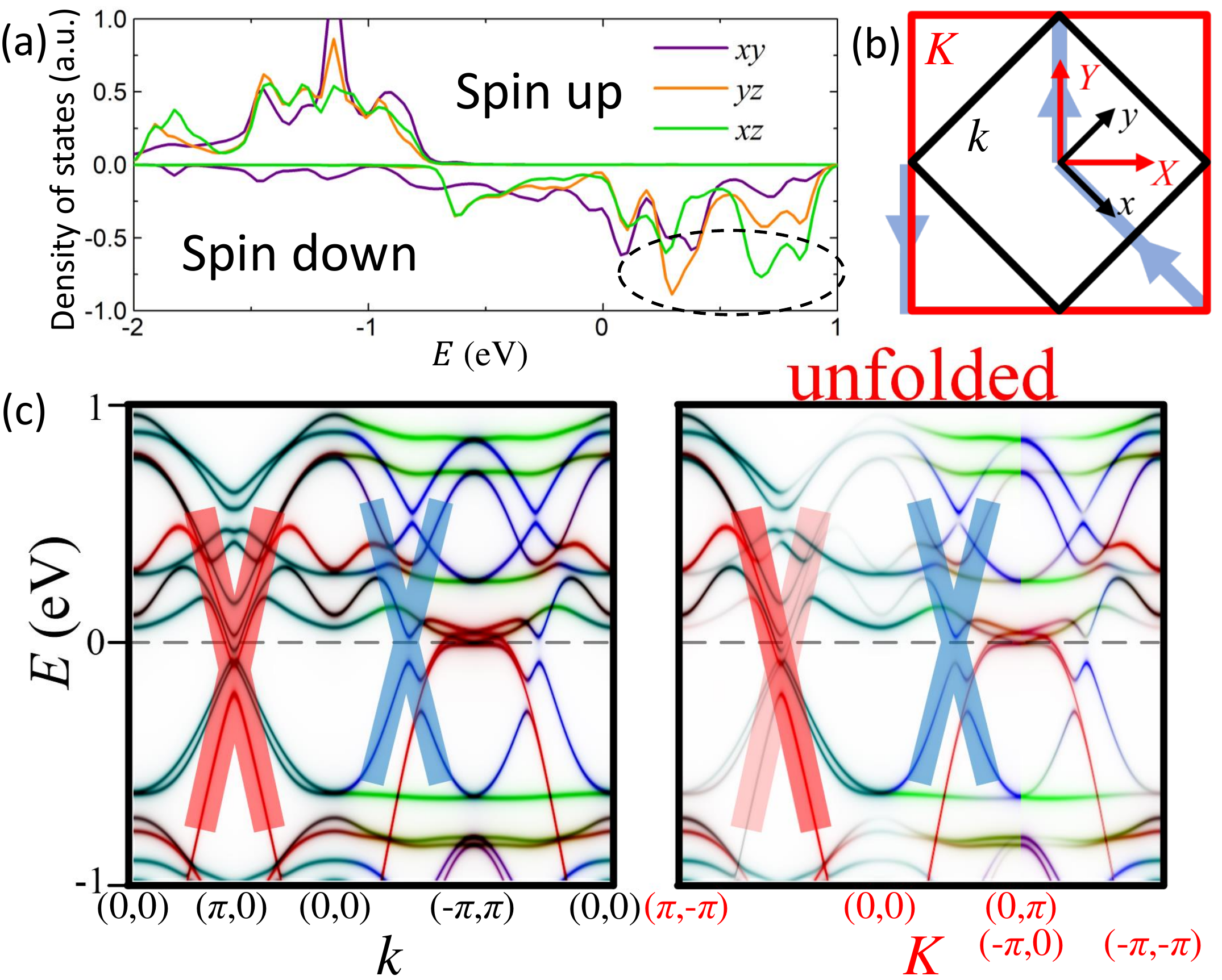}
\end{center}
\vspace{-0.8cm}
\caption{
\label{fig:fig2}
(color online) Electronic structure of the SOO phase.
(a) Local density of states showing orbital order $xz\ne yz$ in down spins.
(b) Illustration for the normal and unfolded Brillouin zones used in (c). 
(c) Normal (left) and unfoled (right) band structure showing the gap near $(\pi/2,\pi/2)$ is \textit{unrelated} to the SOO-induced band folding.
}
\vspace{-0.3cm}
\end{figure}

Before digging into the lower energy physics, let's first examine the physical effects of the SOO on the electronic structure via a qualitative DFT calculation, using the 292K crystal structure~\cite{paper29} and the LDA+$U$ approximation with $U_{eff}=3.4$eV~\cite{paper17}, to improve the LDA results~\cite{paper25}.
As expected, the density of states in Fig.~\ref{fig:fig2}(a) consists of an asymmetric $xz$ and $yz$ contributions in spin down channel, reflecting the SOO.

The most important consequence of the SOO is the band folding and associate gap opening near ($\pi$,0) and (0,$\pi$) highlighted by the transparent red cross in Fig.~\ref{fig:fig2}(c).
This is in essence the main reason for the creation of the small electron and hole pockets observed in the Fermi level~\cite{paper3}.
The origin of this folding and gap open can be illustrated easily by unfolding the band structure to a larger Brillouin zone~\cite{unfolding} shown in Fig.~\ref{fig:fig2}(c), corresponding to a smaller, more symmetric unit cell without SOO.
The unfolded band structure has a diminishing intensity for the bands in one side of the red cross, indicating that they are folded here by the SOO and the associate gap openings near the cross are results of the SOO.

In comparison, the gap opening near ($\pi/2$,$\pi/2$) (marked by the blue cross in Fig.~\ref{fig:fig2}(c)) does not show such a strong contrast in intensity, indicating that this gap is unrelated to the SOO.
Note that this gap is exactly the one observed by previous ARPES~\cite{paper3} and led to the speculation of a finite-momentum order.
Our detailed analysis shows that this gap actually originates from inter-layer coupling of the bilayer, instead~\cite{supp}.
Since this gap originates from effective covalent bonding, it is unrelated to a low-temperature broken symmetry phase and thus would not close at any regular temperature, contrary to the previous speculation.

The above analysis [c.f.~\ref{fig:fig1}(i)] indicates that both SOO and FEO orders occur at much higher temperature to account for the puzzling 48K transition.
On the other hand, an antiferromagnetic order (AFMO) takes place at 56K~\cite{afmpattern}, slightly higher than the 48K transition, suggesting an intimate relation to the latter.

\begin{figure}[th]
\begin{center}
\resizebox{1.0\columnwidth}{!}{\includegraphics{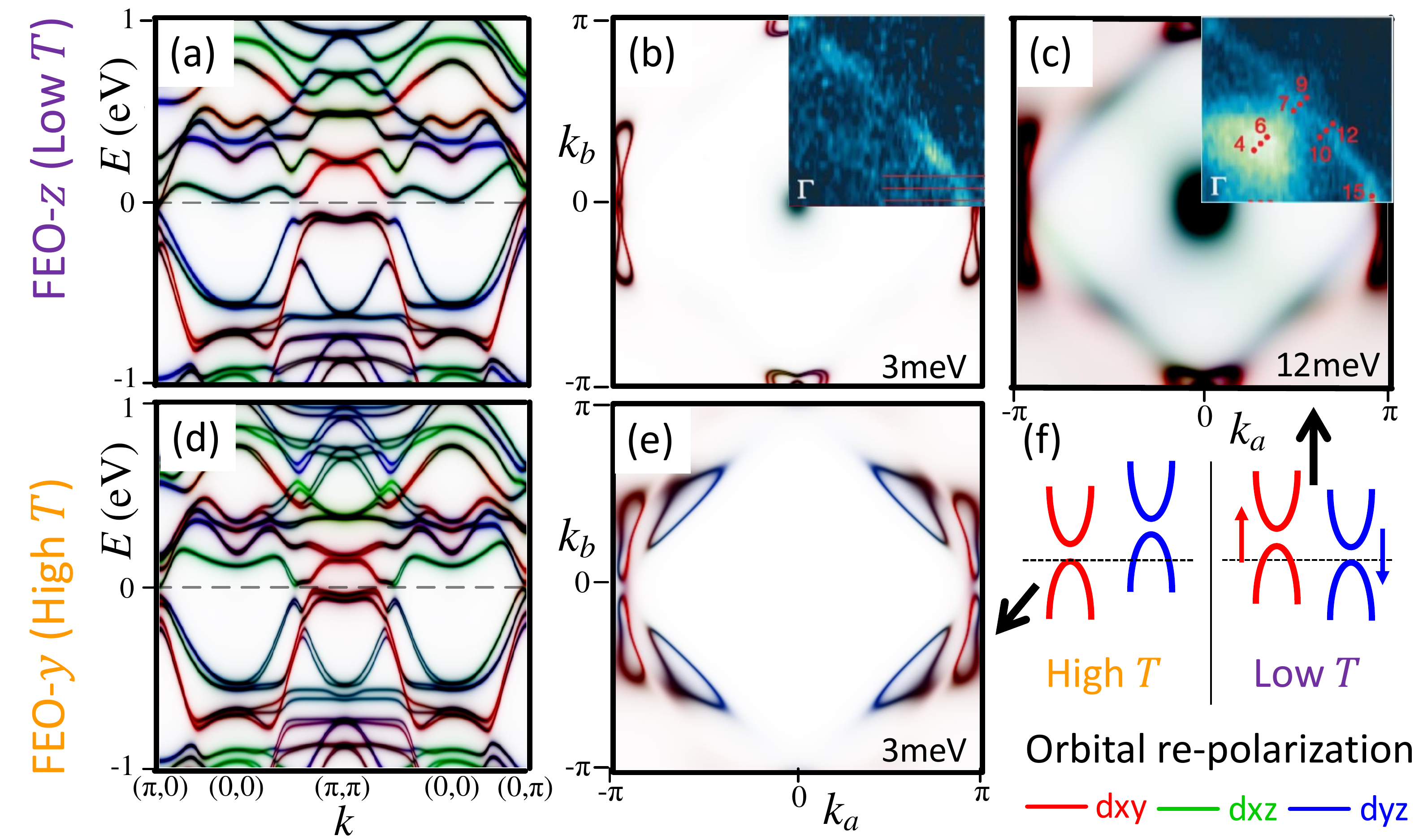}}
\end{center}
\vspace{-0.8cm}
\caption{
\label{fig:fig3}
(color online)
(a)(d) Band structures and (b)(e) Fermi surfaces ($\pm 3$meV) of two distinct electronic states.
(c) States $\pm 12$meV around the Fermi energy, showing a diamond-shaped small gap.
Insets in (b) and (c) show experimental spectra~\cite{paper3} containing additional broadening from instrumental resolution ($7.5$meV) and impurity scattering.
(f) Illustration of the orbital re-polarization and its consequence on the gap energy near ($\pi/2$,$\pi/2$).
}
\vspace{-0.3cm}
\end{figure}

To investigate the 48K first-order phase transition, we proceed by performing DFT calculation using the 40K crystal structure~\cite{paper29}, including AFMO and spin-orbital coupling along the direction of (010) seen in the low-temperature experiment~\cite{afmpattern}.
Interestingly, under the SOO and FEO, we found \textit{two} stable DFT solutions with the same AFMO.
The first state, termed FEO-$z$, has Fermi surface [c.f.~\ref{fig:fig3}(b)] resembling well the experimental one of the low-temperature state~\cite{paper3}, with tiny electron pockets around $(\pi,0)$ and hole pockets around $(0,\pi)$~\cite{paper28}.
It also has diamond-shaped gapped excitations in momentum [c.f.~\ref{fig:fig3}(c)], in excellent agreement with the experimental observation~\cite{paper3}.
Therefore, we conclude that FEO-$z$ state represent well the state below 48K.

The second solution, termed FEO-$y$, has nearly the same total energy as FEO-$z$, but is notably different in electronic structure.
Its Fermi surface shown in Fig.~\ref{fig:fig3}(e) consists of larger pockets, compared to the FEO-$z$ state in Fig.~\ref{fig:fig3}(b).
These larger Fermi pockets naturally indicate higher carrier density and should give a lower resistivity, similar to the state above 48K~\cite{paper28}.
As to be demonstrated in detail below, the difference in the electronic structure between these two states reproduces all the characteristics of the observed changes in the experimental lattice structure from around 56K to slightly below 48K, suggesting strongly that this FEO-$y$ state describe the state around 56K well.

One key microscopic difference between these two states is a notable difference in the occupation of their Ru-$d_{t2g}$ orbitals.
More than $10\%$ of electrons in the FEO-$y$ state are transfered from the $d_{xy}$ orbital to the $d_{yz}$ orbital in the (low-temperature) FEO-$z$ state.
The increased $d_{yz}$ occupation would naturally elongate the local Ru-O octahedra along the $z$-axis, thus compelling a larger tilting of the octahedra, exactly like the experimentally observed changes starting at the 56K magnetic order in Fig.~\ref{fig:fig1}(b)(e).

In other words, the magnetic order is accompanied by a $d_{xy}$ to $d_{yz}$ orbital re-polarization.
At lower temperature, the in-plane ferromagnetic ordering gradually promotes a larger kinetic energy (larger band width) of the intra-layer orbital $d_{xy}$, so the top of its band moves pass the top of the $d_{yz}$ band [c.f.: Fig.~\ref{fig:fig3}(f)], leading to the charge transfer.
Correspondingly, Fig.~\ref{fig:fig3}(b)\&(e) show that the (blue) $d_{yz}$ holes near ($\pi/2$,$\pi/2$) in the FEO-$y$ state are transfered to the (red) $d_{xy}$ holes near (0,$\pi$) in the FEO-$z$ state.
As illustrated in Fig.~\ref{fig:fig3}(f), this also reveals a clean gap around the chemical potential near ($\pi/2$,$\pi/2$), and thus gives the impression that an "ordering gap" is opened in the low temperature phase.

\begin{figure}[th]
\begin{center}
\includegraphics[width=\columnwidth,clip,angle=0]{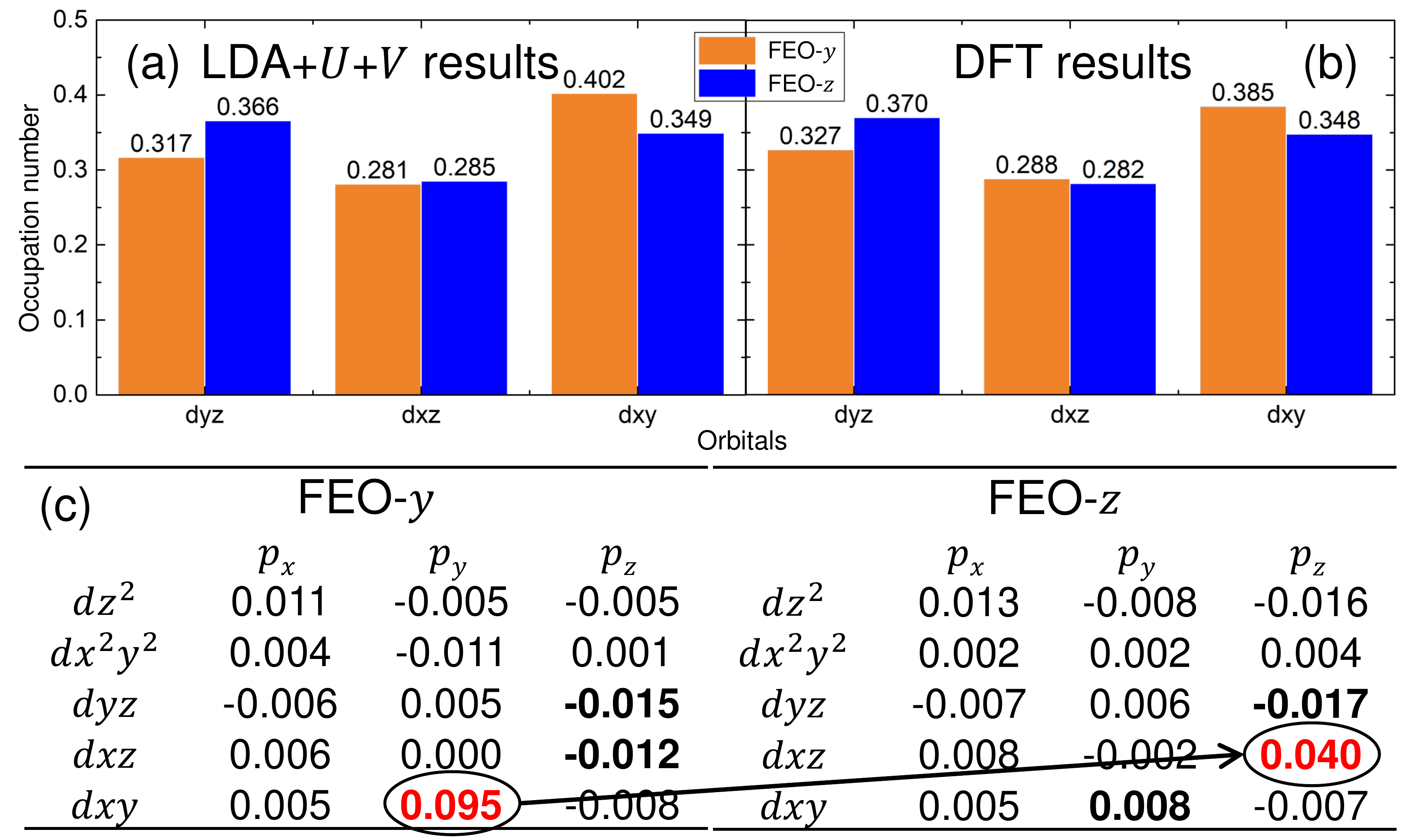}
\end{center}
\vspace{-0.8cm}
\caption{
\label{fig:fig5}
(color online) Comparison of orbital occupation of FEO-$y$ and FEO-$z$ states from (a) LDA+$U$+$V$ and (b) DFT calculation.
(c) FEO order parameters $\langle d^{\dagger}p\rangle$ from LDA+$U$+$V$.
}
\vspace{-0.5cm}
\end{figure}

Now, to better incorporate the FEO, a theory allowing a spontaneous breaking of the local parity symmetry is necessary.
To this end, we implement a LDA+$U$+$V$ method that includes additional Coulomb interactions between the Ru $d$- and $p$-orbitals:
\begin{eqnarray}
\label{eq:eqn1}
&&\hat{H}=\hat{H}_{LDA+U}+\hat{H}_{U}-\hat{H}_{U}^{0}+\hat{H}_{V}-\hat{H}_{V}^{dc}\nonumber\\
&&\hat{H}_{U}=\sum _{{\rm Ru}}\sum _{d,d^\prime,\sigma,\sigma^\prime}Uc^{\dagger}_{d\sigma} c^{\dagger}_{d^\prime\sigma^\prime}c_{d^\prime\sigma^\prime}c_{d\sigma}\nonumber\\
&&\hat{H}_{V}=\sum _{{\rm Ru}}\sum _{d,p,\sigma,\sigma^\prime}Vc^{\dagger}_{d\sigma} c^{\dagger}_{p\sigma^\prime}c_{p\sigma^\prime}c_{d\sigma}+V^\prime c^{\dagger}_{p\sigma} c^{\dagger}_{d\sigma^\prime}c_{p\sigma^\prime}c_{d\sigma}\nonumber\\
&&\hat{H}_{V}^{dc}=\sum _{{\rm Ru}}\sum _{d,p,\sigma}\left[(V-V^\prime)n_{p\sigma} n_{d\sigma}+Vn_{p\sigma} n_{d-\sigma}\right]
\end{eqnarray}
where $\hat{H}_{LDA+U}$ is extracted from LDA+$U$ calculation, covering bands $\sim\pm5$eV around the chemical potential,  using symmetric Wannier functions~\cite{dftwannier} of Ru-$d$, Ru-$p$, and O-$p$ symmetries.
$\hat{H}_{U}$ contains the local Coulomb interaction between Ru-$d$ orbitals, and $\hat{H}_{V}$ between Ru-$d$ and Ru-$p$ orbitals.
$\hat{H}_{U}^{0}$ is the initial $\hat{H}_{U}$ from LDA+$U$. 
$\hat{H}_{V}^{dc}$ is the double counting term for $\hat{H}_{V}$, treated similarly to the SIC method~\cite{anisimov1993density} in LDA+$U$.
The indexes $d,d^\prime$ and $p$ denotes the $d$- and $p$-orbitals, and $\sigma,\sigma^\prime$ the spin of electrons.

We then solve the Hamiltonian with the self-consistent Hartree-Fock approximation~\cite{supp}, treating the full 16$\times$16 density matrix, $\langle c^{\dagger}_{m\sigma} c_{m^\prime\sigma^\prime} \rangle$, as the mean-field.
(Here $m$ refers to the $p$- or $d$-orbitals.)
We choose a reasonable choice of $U\sim 3.4$eV~\cite{paper17}, and $V-V^\prime\sim V\sim 3$eV to illustrate the qualitative trends of the physics.

Similar to the DFT calculation, the self-consistent iteration also converges to two different electronic states.
The same (but even stronger) difference in the occupation of $d_{xy}$ and $d_{yz}$ orbitals is found in Fig.~\ref{fig:fig5}(a)(b).
This means that the model is not only realistic enough to capture the above orbital re-polarization effect, but also improves this particular effect over the DFT calculation, owing to the additional $p$-$d$ repulsion $V$.

\begin{figure}[th]
\begin{center}
\includegraphics[width=\columnwidth,clip,angle=0]{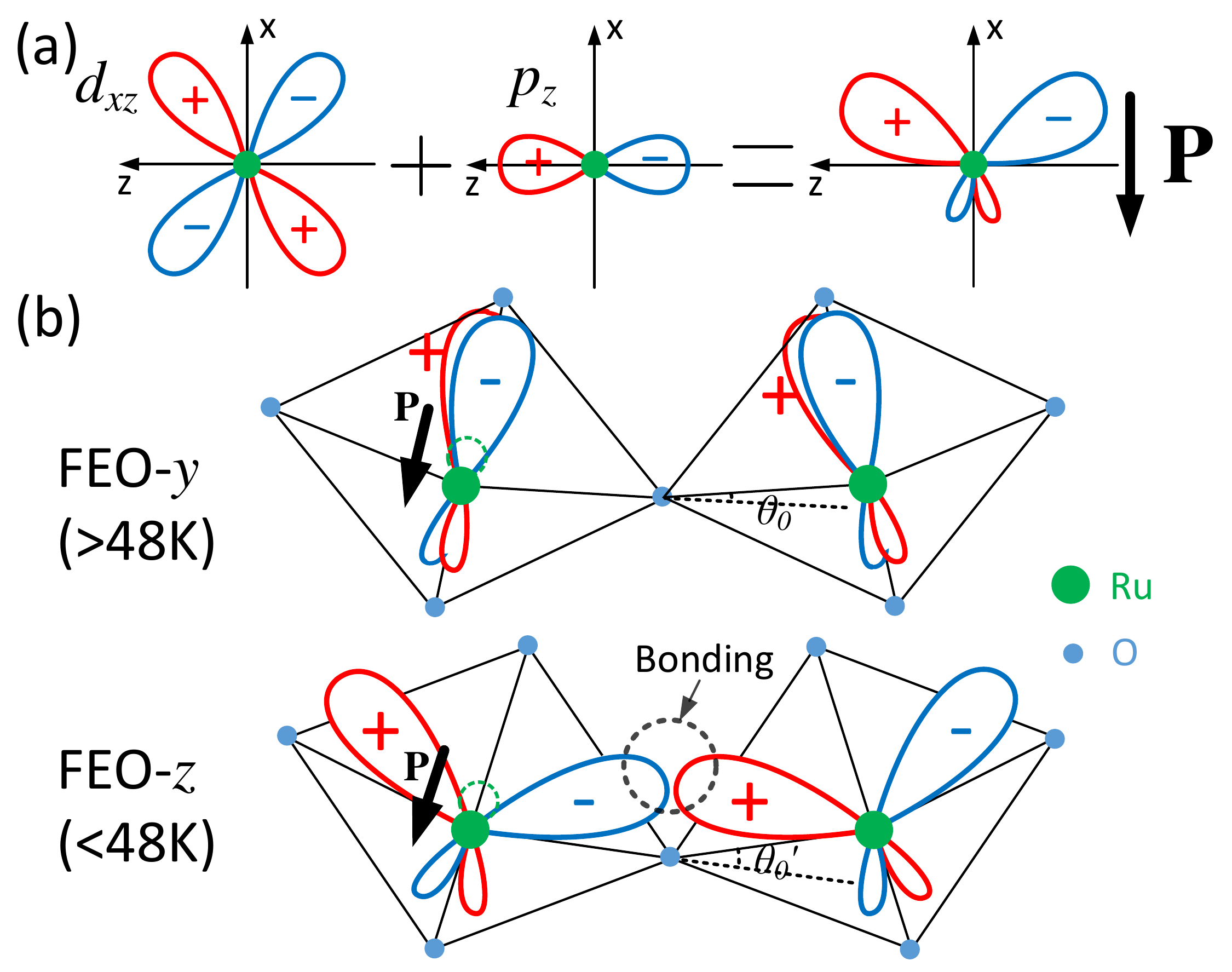}
\end{center}
\vspace{-0.8cm}
\caption{
\label{fig:fig4}
(color online) (a) Illustration of charge polarization and lobe bending due to $d$-$p$ hybridization  $\langle d^{\dagger}p\rangle$
(b) Illustration of bi-layer bond formation, strengthened octahedral tilting, and the switching of FEO order from $d_{xy}+ p_{y}$ to $d_{xz}+ p_{z}$.
}
\vspace{-0.3cm}
\end{figure}

Notice that other than the standard orbital- and magnetic-order parameters, the density matrix also includes the many-body interaction induced $p$-$d$ hybridization, $\langle c^{\dagger}_{d\sigma} c_{p\sigma} \rangle$.
These hybridizations can be regarded as (spontaneous symmetry breaking) local order parameters of the FEO~\cite{liu1988electron,portengen1996linear,batista2002electronic,batista2007electronic}, since they couple directly to the broken local parity symmetry and is related to the local electric dipole.
This is illustrated in Fig.~\ref{fig:fig4}(a), in which an odd $x$-parity $d_{xz}$ orbital hybridizes with an even $x$-parity $p_z$ orbital.
The resulting orbital thus breaks the $x$-parity and develops an local electric dipole, \textbf{P}, and pushes the Ru along the $-x$ direction away from the center of the octahedra [c.f.~\ref{fig:fig1}(a)(c)].

Figure~\ref{fig:fig4}(a) also shows that hybridization with the $p_z$ orbital bends the $d_{xz}$ orbital toward the $z$-direction, and thus tends to elongate the octahedron along the $z$-direction, perpendicular to the direction of the local electric dipole.
This is in excellent agreement with the experimentally observed perpendicular directions of the octahedral elongation and the deviation of the central Ru in the FEO phase illustrated in Fig.~\ref{fig:fig1}(h).

Interestingly, by examining these FEO order parameters in Fig.~\ref{fig:fig5}(c), one finds that the FEO-$y$ state (above 48K) is characterized by the $d_{xy}+p_y$ hybridization, while the FEO-$z$ state (below 48K) by $d_{xz}+p_z$, as illustrated in Fig.~\ref{fig:fig4}(b). 
That is, these two states actually host different FEO orders, even though they both break the local $x$-parity symmetry.
Such a switching of FEO is naturally reflected in the change of octahedral O-Ru-O deformation given in Fig.~\ref{fig:fig1}(d): $\theta _{z}$ that couples to the $xz+z$ order parameter [c.f.: Fig.~\ref{fig:fig1}(a)], demonstrates an increase below the 56K magnetic transition and continue across the 48K transition, while the $\theta_{y}$ experiences the opposite.
Even more, the strong reduction of the ferroelectric order parameter (by roughly one half), from 0.095 to 0.04, agrees quite well with the experimental observation in Fig.~\ref{fig:fig1}(c).

These two effects (elongation of the orbital and the switching of order parameters) reveal the microscopic nature of the 48K first-order transition.
Below the 56K magnetic transition, the above mentioned orbital re-polarization gradually enhances the tilting of octahedra (from $\theta _{0}$ to $\theta _{0}^\prime$ in Fig.~\ref{fig:fig4}(b)).
Meanwhile, the stronger "$xz+z$" order elongates and bends the orbital toward the $z$-axis, to the point that this orbital between the bi-layer starts to bind chemically.
In turn, this newly developed bond further enhances the tilting, which promotes the switching of FEO from $xy+y$ to $xz+z$ one, thus enhancing the elongation and the bending of $xz+z$ orbital.
This self-consistent feedback gives rise to an abrupt first-order transition.
This process naturally gives rise to the experimentally observed first-order jump in the octahedral tilting [c.f.~\ref{fig:fig1}(e)] and the associated collapse in $c$-axis~\cite{paper29} lattice constant (due to stronger inter-layer bonding.)

Unlike a general bond-forming first-order phase transition observed in some materials under high pressure~\cite{RN29,PhysRevB.89.121109}, as far as we can tell this case might be the first example with a (temperature-) magnetism-induced switching from one FEO to another of the same polarization direction.
Interestingly, the associate orbital repolarization shifts the gap below chemical potential, producing an appearance of a gap opening ordering (which is typically a second-order phase transition).
This example illustrates beautifully the interesting compexity of strongly correlated materials with multiple degrees of feedoms and the rich interplay between various ordering in such systems.

In summary, based on our arnalysis of structural data, DFT calculation, and LDA+$U$+$V$ calculation, we provide a microscopic picture of all the electronic phases of Ca$_3$Ru$_2$O$_7$.
Particularly, the outstanding issue of the 48K first-order phase transition is explained as an inter-layer bond forming transition through a magnetism-induced orbital-reporlarization.
Interestingly, the transition is accompanied by a switching of two distinct FEOs of different order parameters.
Our picture naturally account for the rich interplay among metallic FEO, AFM order, SOO and lattice structure observed previously~\cite{paper20,paper28,paper6,paper21,paper24,paper10}.
Our LDA+$U$+$V$ method should be generally applicable to the study of other strongly correlated multi-ferroic materials.
Our study not only resolves a long-standing puzzle of this phase transition in this material, but also demonstrates perhaps the first example of transition between multiple emergent ferroelectric orders in bad metals, resulting from interplay between multi-ferroic orders.

We thank Sudeshna Sen, Hui Xing, Ying Liu, and H. Padmanabhan for useful discussions.  Work supported by National Natural Science Foundation of China \#11674220 and 11447601, and Ministry of Science and Technology \#2016YFA0300500 and 2016YFA0300501.

\bibliography{Ti}

\end{document}